\documentclass[preprint,showpacs,preprintnumbers,amsmath,amssymb]{revtex4}

\newcommand{\ds}{\displaystyle}

\usepackage{graphicx}
\usepackage{dcolumn}
\usepackage{bm}

\newcommand{\be}{\begin{equation}}
\newcommand{\en}{\end{equation}}
\newcommand{\bea}{\begin{eqnarray}}
\newcommand{\ena}{\end{eqnarray}}

\begin{document}


\title{Emergent universe in a Jordan-Brans-Dicke theory}

\author{Sergio del Campo}
 \email{sdelcamp@ucv.cl}
\affiliation{ Instituto de F\'{\i}sica, Pontificia Universidad
Cat\'{o}lica de Valpara\'{\i}so, Casilla 4059, Valpara\'{\i}so,
Chile.}
\author{Ram\'on Herrera}
\email{ramon.herrera@ucv.cl} \affiliation{ Instituto de
F\'{\i}sica, Pontificia Universidad Cat\'{o}lica de
Valpara\'{\i}so, Casilla 4059, Valpara\'{\i}so, Chile.}
\author{Pedro Labra\~{n}a}
 \email{pedro.labrana@ucv.cl}
\affiliation{ Instituto de F\'{\i}sica, Pontificia Universidad
Cat\'{o}lica de Valpara\'{\i}so, Casilla 4059, Valpara\'{\i}so,
Chile.}

\date{\today}

\begin{abstract}

 In this paper we study emergent universe model  in the context of a
 self interacting Jordan-Brans-Dicke theory. The model presents a stable past
 eternal static solution which eventually enters a phase where the
 stability of this solution is broken leading to an inflationary
 period. We also establish constraints for the different parameters appearing in our
 model.
\end{abstract}

\maketitle

\section{Introduction}

Cosmological inflation has become an integral part of the standard
model of the universe. Apart from being capable of removing the
shortcomings of the standard cosmology, it gives important clues
for large scale structure formation. The scheme of inflation
\cite{Guth, Albrecht, Linde1, Linde2} (see \cite{libro} for a
review) is based on the idea that there was a early phase in which
the universe evolved through accelerated expansion in a short
period of time at high energy scales. During this phase, the
universe was dominated by a potential $U(\Psi)$ of a scalar field
$\Psi$, which is called the inflaton.

 Singularity theorems have been devised that apply in the
inflationary context, showing that the universe necessarily had a
beginning (according to classical and semi-classical theory)
\cite{Borde:1993xh,Borde:1997pp, Borde:2001nh,
Guth:1999rh,Vilenkin:2002ev}. In other words, according to these
theorems, the quantum gravity era cannot be avoided in the past
even if inflation takes place.
However, recently, models that escape this conclusion has been
studied in Refs.
\cite{Ellis:2002we,Ellis:2003qz,Mulryne:2005ef,Mukherjee:2005zt,
Mukherjee:2006ds,Banerjee:2007qi,Nunes:2005ra,Lidsey:2006md}.
These models do not satisfy the geometrical assumptions of these
theorems. Specifically, the theorems assume that either {\bf i)}
the universe has open space sections, implying $k = 0$ or $-1$, or
\textbf{ii)} the Hubble expansion rate $H = \dot{a}/a$ is bounded
away from zero in the past, $H > 0$, where $a$ is the scale
factor.
In particular, Refs.
\cite{Ellis:2002we,Ellis:2003qz,Mulryne:2005ef,Mukherjee:2005zt,
Mukherjee:2006ds,Banerjee:2007qi,Nunes:2005ra,Lidsey:2006md}
consider closed models in which k = +1 and H can become zero, so
that both assumptions \textbf{i)} and \textbf{ii)} of the
inflationary singularity theorems are violated. The models studied
in Refs. \cite{Ellis:2002we,Ellis:2003qz} obey general relativity,
contain only ordinary matter, and (minimally coupled) scalar
fields.
In these models, the universe is positively curved and initially
in a past eternal classical Einstein static (ES) state that
eventually evolves into a subsequent inflationary phase.
Such models are appealing since they provide specific examples of
non–singular (geodesically complete) inflationary universes.
Furthermore, it has been proposed that entropy considerations
favor the ES state as the initial state for our universe
\cite{Gibbons:1987jt,Gibbons:1988bm}.

However, the models based on general relativity with ordinary
matter suffer from a number of important shortcomings. In
particular, the instability of the ES state (represented by a
saddle equilibrium point in the phase space of the system, see
\cite{Mulryne:2005ef,Banerjee:2007qi,Nunes:2005ra,Lidsey:2006md})
makes it extremely difficult to maintain such a state for an
infinitely long time in the presence of fluctuations, such as
quantum fluctuations, that will inevitably arise.
As in the emergent universe scenario, it is assumed that the
initial conditions are specified such that the static
configuration represents the past eternal state of the universe,
out of which the universe slowly evolves into an inflationary
phase. The instability of the ES solution ensures that any
perturbations, no matter how small, rapidly force the universe
away from the static state, thereby aborting the scenario.
Some models have been proposed to solve the stability problem of
the asymptotic static solution. They consider non-perturbative
quantum corrections of the Einstein field equations, either coming
from a 'semiclassical' state in the framework of loop quantum
gravity (LQG) \cite{Mulryne:2005ef,Nunes:2005ra} or braneworld
cosmology with a timelike extra dimension \cite{Lidsey:2006md,
Banerjee:2007qi}. Other possibilities to consider are the
Starobinsky model or exotic matter \cite{Mukherjee:2005zt,
Mukherjee:2006ds}.

The Jordan-Brans-Dicke (JBD) \cite{Jbd} theory is a class of
models in which the effective gravitational coupling evolves with
time. The strength of this coupling is determined by a scalar
field, the so-called Brans-Dicke field, which tends to the value
$G^{-1}$, the inverse of the Newton's constant. The origin of
Brans-Dicke theory is in Mach's principle according to which the
property of inertia of material bodies arises from their
interactions with the matter distributed in the universe. In
modern context, Brans-Dicke theory appears naturally in
supergravity models, Kaluza-Klein theories and in all the known
effective string actions \cite{Freund:1982pg, Appelquist:1987nr,
Fradkin:1984pq, Fradkin:1985ys, Callan:1985ia, Callan:1986jb,
Green:1987sp}.

In this work we consider a JBD  model and determine whether such a
model could fit the general characteristics of an emergent
universe scenario: a stable static past asymptotic solution
followed by a period of de Sitter inflation.
Here, instead of employing a time-like extra dimension or
examining a past eternal static solution that lies in the
semiclassical quantum gravity regimen of the theory, we work more
conventionally, keeping our model just at the classical level, in
the spirit of Refs.~\cite{Ellis:2002we,Ellis:2003qz}.

The paper is organized as follows. In Sect.~\ref{secti} we review
briefly the cosmological equations of the JBD model. In
Sect.~\ref{static} the existence and nature of a static solution
is discussed. In Sect.~\ref{inf} we study the dynamics that lead
to the emergence of an inflationary universe. In Sect.~\ref{model}
we present a specific model that satisfies the requirements of an
emergent universe in the scheme of JBD theories. In
Sect.~\ref{conc} we summarize our results.

\section{The Model \label{secti}}

We consider the following JBD action \cite{Jbd} for a
self-interacting potential and matter, given by

 \begin{eqnarray}
 \label{ac1}
\ds S =  \int{d^{4}x\,\sqrt{-g}}\bigg[\frac{1}{2}\,\,\Phi\,R\,
-\,\frac{1}{2}\,\frac{w}{\Phi}\,\partial_{\mu}\Phi
\,\partial^{\mu}\Phi -\, V(\Phi) +{\cal{L}}(\Psi)\bigg],
 \end{eqnarray}
where ${\cal{L}}(\Psi)$ denote the Lagrangian density of the
matter
$${\cal{L}}(\Psi)\,=\,\frac{1}{2}\partial_{\mu}\Psi
\partial^{\mu}\Psi\,-\, U(\Psi),$$
$R$ is the Ricci scalar curvature, $\Phi$ is the JBD scalar field,
$w$ is the JBD parameter and $V(\Phi)=V$ is the potential
associated to the  field $\Phi$. Here $\Psi$ is the standard
inflaton field and $U(\Psi)$ its effective potential. In this
theory $1/\Phi$ plays the role of the gravitational constant,
which changes with time. This action also matches the low energy
string action for $w=-1$ \cite{Green:1987sp}.

The Friedmann-Robertson-Walker metric is described by
 \be \ds
d{s}^{2}\,=\, d{t}^{2}\,-\, a(t)^{2}\, \,\, d\Omega^{2}_{k}\,\,,
\label{met} \en where $a(t)$ is the scale factor, $t$ represents
the cosmic time and $d\Omega^{2}_{k}$ is the spatial line element
corresponding to the hypersurfaces of homogeneity, which could
represent a three-sphere, a three-plane or a three-hyperboloid,
with values $k$=1, 0, -1, respectively. From now on,  we will
restrict ourselves to the case $k$ = 1.

Using the metric~(\ref{met}), with $k$=1, in the
action~(\ref{ac1}), we obtain the following field equations:

\begin{equation}
 H^2\,+\frac{1}{a^2}\,+H\frac{\dot{\Phi}}{\Phi}=\frac{\rho}{3\,\Phi}+
 \frac{w}{6}\left(\frac{\dot{\Phi}}{\Phi}\right)^2+\frac{V}{3\,\Phi}\label{key_02},
\end{equation}

\begin{equation}
2\frac{\ddot{a}}{a}+H^2\,+\frac{1}{a^2}\,+\frac{\ddot{\Phi}}{\Phi}+2\,H\,\frac{\dot{\Phi}}{\Phi}+
 \frac{w}{2}\left(\frac{\dot{\Phi}}{\Phi}\right)^2-\frac{V}{\Phi} =-\frac{P}{\Phi}\label{dda},
\end{equation}
\begin{equation}
\ddot{\Phi}+3H\dot{\Phi}=\frac{(\rho-3
P)}{(2w+3)}+\,\frac{2}{2w+3}\left[2\,V-\Phi\,V'\right]\;,
\label{key_01}
\end{equation}
and the conservation of energy-momentum implies that
\begin{equation}
\dot{\rho}+3H(\rho+P)=0,
\end{equation}
 or equivalently

\begin{equation} \ddot{\Psi}+3H\dot{\Psi}=-\frac{\partial
U(\psi)}{\partial\Psi},\label{3}
\end{equation}
where $V'=dV(\Phi)/d\Phi$. Dots mean derivatives with respect to
time, units are such that $8\pi\,G=1$ and $c=\hbar=1$.

Here the energy density $\rho$ and pressure $p$ are given by

\begin{equation}
\rho = \frac{\dot{\Psi}^2}{2}+U(\Psi),
\end{equation}
and
\begin{equation} P = \frac{\dot{\Psi}^2}{2}-U(\Psi)\,.
\end{equation}

We could write an effective state equation for this scalar field
given by $P=(\gamma -1)\,\rho$, where the equation of state
"parameter", $\gamma$, is defined by

\begin{equation} \label{v1g}
\gamma=2\left(1-\frac{U(\Psi)}{\rho}\right).
\end{equation}

In the situation where the scalar potential of the inflaton field
is constant, the state parameter is a function only of the scale
factor.
%

\section{Static universe \label{static}}

\subsection{Static solution in JBD model\label{static1}}

In the JBD model the  static universe is characterized by the
conditions $a=a_0=Const.$, $\dot{a}_0=0=\ddot{a}_0$ and
$\Phi=\Phi_0=Cte.$, $\dot{\Phi}_0=0=\ddot{\Phi}_0$. Following the
same scheme as the static Einstein model, we are going to consider
that the matter potential $U(\Psi)$ becomes asymptotically flat in
the limit $\Psi \rightarrow -\infty$, that is
$U(\Psi)=U_0=Const.$. In this limit, the initial conditions are
specified such that the static configuration represents the past
eternal state of the universe, out of which the universe slowly
evolves into an inflationary phase. Then from equations
(\ref{key_02}) to (\ref{key_01}) and using the equation of state
 $P=(\gamma -1)\,\rho$, we obtain the following equations
of stability

\begin{eqnarray}
0 &=& \frac{\rho_0}{3\,\Phi_0} + \frac{V_0}{3\,\Phi_0} -
\frac{1}{a^2_0}\,, \nonumber\\
\nonumber \\
0 &=& (4-3\,\gamma_0)\,\rho_0 + 4V_0 - 2\Phi_0\,V'_0\,, \label{eqst}\\
\nonumber\\
0 &=& \frac{1}{a^2_0} + (\gamma_0 -1)\,\frac{\rho_0}{\Phi_0} -
\frac{V_0}{\Phi_0}\,, \nonumber
\end{eqnarray}
where  $V_0=V(\Phi_0)$,
\mbox{$V_0'=(dV(\Phi)/d\Phi)_{\Phi=\Phi_0}$} and
\mbox{$\gamma_0=2\left(1-\frac{U(\Psi_0)}{\rho_0}\right)$}.


The equations (\ref{eqst}) are satisfied if the following
conditions are fulfilled

\begin{eqnarray}
\label{v1g} \gamma_0 &=& 2\,\frac{\Phi_0}{a_0^2\,\rho_0} =
2\left(1-\frac{U_0}{\rho_0}\right),\\
\nonumber \\
\label{v2} a_0^2 &=& \frac{3}{V'_{0}}\,, \\
\nonumber \\
\label{v} \rho_0 &=& V'_{0}\,\Phi_0 -\,V_0.
\end{eqnarray}

We can obtain the velocity at which the scalar field $\Psi$ is
rolling along the constant potential $U_0$ as a function of the
static values of the scale factor $a_0$, Brans-Dicke field
$\Phi_0$ and energy density $\rho_0$

\begin{equation}
\label{vel} \dot{\Psi}^2_0=2\,\frac{\Phi_0}{a_0^2}\,.
\end{equation}

Note that in order to obtain a static solution we need to have a
non-zero JBD potential with a non-vanishing derivative at the
static point $\Phi=\Phi_0$. The original Brans-Dicke model
corresponds to $V(\Phi)=0$. However, non-zero $V(\Phi)$ is better
motivated and appears in many particle physics models. In
particular, $V(\Phi)$ can be chosen in such a way that $\Phi$ is
forced to settle down to a non-zero expectation value, $\Phi
\rightarrow m_p^2$, where $m_p = 10^{19} GeV$ is the value of the
Planck mass today. On the other hand, if $V(\Phi)$ fixes the field
$\Phi$ to a non-zero value, then time-delay experiments place no
constraints on the Brans-Dicke parameter $w$ \cite{La:1989pn}.

\subsection{Oscillations \label{staticB}}

As we have mentioned above, one important point that we have to
determine is whether the static JBD solution found in the previous
section corresponds to an stable solution. In order to see this,
let us consider small perturbation about the static solution for
the scale factor and the JBD field. We set

\begin{equation} \label{s1}
a(t)=a_0\left[1+\varepsilon(t)\right],
\end{equation}
and
\begin{equation} \label{s2}
\Phi(t)=\Phi_0\,[1+\beta(t))]\,,
\end{equation}
where $\varepsilon\ll 1$ and $\beta\ll 1$ are small perturbations.
By introducing the expressions (\ref{s1}) and (\ref{s2}) into
Eq.~(\ref{dda}) and Eq.~(\ref{key_01}) and retaining only at the
linear order in $\epsilon$ and $\beta$ we obtain the following
coupled equations:

\begin{equation} \label{s3}
\ddot{\varepsilon}
-\frac{4}{a_0^2}\,\varepsilon+\frac{\ddot{\beta}}{2}-\frac{\beta}{a_0^2}=0\,,
\end{equation}
and
\begin{equation} \label{s4}
\ddot{\beta}-\frac{1}{3+2w}\left[\frac{12}{a_0^2}\,
\varepsilon+\left(\frac{6}{a_0^2}-2\Phi_0\,V_0''\right)\,\beta\right]=0\,,
\end{equation}

where $V_0''=(d^2V(\Phi)/d\Phi^2)_{\Phi=\Phi_0}$.

Here, we have used that
\begin{equation} \label{p1}
\rho=\rho_0+\delta\rho(\varepsilon)\approx\rho_0-3\gamma_0\,\rho_0\,\varepsilon
\,,
\end{equation}
and
\begin{equation} \label{p2}
\gamma=\gamma_0+\delta\gamma(\varepsilon)\approx
\gamma_0-6\frac{\gamma_0\,U_0}{\rho_0}\,\varepsilon \,.
\end{equation}


From the system of Eqs.(\ref{s3}) and (\ref{s4}) we can obtain the
frequencies of small oscillation


\[
\hspace{-6.0cm}\omega_{\pm}^2 =
\frac{1}{a_0^2(3+2w)}\bigg[a_0^2\Phi_0\,V_0''-2(2+3w)\]
\begin{equation} \label{pc2}
\hspace{4.0cm}\pm \,
\sqrt{[a_0^2\Phi_0V_0'']^2+4a_0^2\Phi_0V_0''(3+2w)+8w(3+2w)}\,\bigg].
\end{equation}

The static solution is stable if $\omega_{\pm}^2>0$. Assuming that
$(3+2w)>0$, we found that the following inequalities must be
satisfied in order to have a stable static solution

\begin{equation} \label{cc1}{\textstyle
0<a_0^2\Phi_0\,V_0''<\frac{3}{2}\,,}
\end{equation}
and
\begin{equation}\label{cc2}{\textstyle
-\frac{3}{2}<w<-\frac{1}{4}\left[\sqrt{9-6a_0^2\Phi_0V_0''}+(3+a_0^2\Phi_0V_0'')\right].}
\end{equation}

These inequalities restrict the parameters of the model. The first
imposes a condition on the JBD potential, specifically for its
first and second derivatives:  $0< V_0'' <V_0'/(2\Phi_0)$. The
second inequality  restrictions  the values of the JBD parameter.
Notice that this inequality imposes that $w<0$. JDB models with
negative values of $w$ have been considered in the context of late
acceleration expansion of the universe \cite{Bertolami:1999dp,
Banerjee:2000mj}, but also appear in low energy limits of string
theory \cite{Green:1987sp}.

In our case we are going to choose the JBD potential in such a way
that $\Phi$ will be forced to stabilize at a constant value
$\Phi_f$ at the end of the inflationary period, see next section.
Then, we can recover General Relativity by setting $\Phi_f =
m_p^2$, therefore whatever we choose for $w$ in our model, it does
not contradict the solar system bound on $w$ \cite{La:1989pn,
Sen:2001ki}.

\section{Leaving the Static Regime  \label{inf}}

The discussion of the previous section determined the behavior of
the model in the case of a constant potential for the scalar field
$\Psi$.
However, any realistic inflationary model clearly requires the
potential to vary as the scalar field evolves. Here, following
Ref.~\cite{Mulryne:2005ef} and with the emergent inflationary
models in mind, we  consider a general class of potentials that
approach  a constant $U_0$ as $\Psi \rightarrow -\infty$ and rise
monotonically once the value of the scalar field exceeds a certain
value.

The overall effect of increasing the potential is to distort the
equilibrium behavior expressed by Eqs.~(\ref{v1g}) and
(\ref{vel}). The inclusion of the derivative term in the equation
of the scalar field $\Psi$ produce changes in its equilibrium
velocity, Eq.~(\ref{vel}), breaking the static solution. In
particular, the field $\Psi$ decelerates as it moves further up
the potential, subsequently reaching a point of maximal
displacement and then rolling back down. If the potential has a
suitable form in this region, slow-roll inflation will occur.

On the other hand, in the slow-roll regimen the scalar potential
evolves slowly. In that case we can consider $U(\Psi)\sim
const.=U_{inf}$. Then, as  mentioned in Ref.~\cite{Green:1996xe},
Eqs. (\ref{key_02}) and (\ref{key_01}) have an exact static
solution for a particular value of $\Phi$, driving a de Sitter
expansion. This occurs when the right hand side of
Eq.~(\ref{key_01}) becomes zero. Denominating this quasi-static
value of the JBD field as $\tilde{\Phi}$, it satisfies the
following condition

\begin{equation}
\label{condinf}
 4U_{inf}+ 4V(\tilde{\Phi}) -
2\tilde{\Phi}\,V'(\tilde{\Phi})=0\,.
\end{equation}

Then, once  the scalar field starts to move in the slow-roll
regime, the JBD field goes to the value $\tilde{\Phi}$ and the
universe begins a de Sitter expansion with

\begin{equation}
H^2 = \frac{1}{3\tilde{\Phi}}\left[U_{inf}+V(\tilde{\Phi})\right].
\end{equation}

For example, the JBD potential $V(\Phi)=\lambda(\Phi^2-\nu)^2$
satisfies this condition, see Ref.~\cite{Green:1996xe}. In the
next section we  consider a specific potential which satisfies
Eq.~(\ref{condinf}) together with the requirements of a static
solution described in Sect.~\ref{static}.

Finally, during the evolution of $\Psi$ over $U(\Psi)$ to zero,
the JBD field evolves slowly to its final value $\Phi_f$, at which
the expression $2V-\phi\,V'$ vanishes. We consider the value
$\Phi_f$ as the current  value of the JBD field.

Also, we can determinate the existence of the inflationary period
by introduce the dimensionless slow-roll parameter

\begin{equation}
\epsilon=-\frac{\dot{H}}{H^2}\simeq\frac{2}{(3+2w)}+\frac{\Phi}{2}
\left(\frac{U_{,\psi}}{V+U}\right)^2
 + \frac{V'\Phi}{(3+2w)(V+U)}\left[\frac{V'\Phi}{V+U}-3\right].
\end{equation}

Then, the inflationary regime takes place if the parameter
$\epsilon$ satisfies  the inequality  $\epsilon<1$, a condition
analogous to the requirement $\ddot{a}> 0$.
We note that if Eq.~(\ref{condinf}) is satisfied (i.e.
$\Phi=\tilde{\Phi}$) and the scalar potential $U(\Psi)$ satisfies
the requirement of an inflationary potential, we get $\epsilon<1$.
In the next Section we study a particular model that follows the
behavior described above.

\section{A SPECIFIC MODEL OF AN EMERGING UNIVERSE}
\label{model}

 From a dynamical point of view, the emergent universe
scenario can be realized in the context of JBD  if the scalar
potential $U(\Psi)$ satisfies a number of weak constraints.
Asymptotically, it should have a horizontal branch as $\Psi
\rightarrow -\infty$ such that $dU/d\Psi \rightarrow 0$ and
increase monotonically in the region $\Psi > \Psi_{grow}$ where
without loss of generality we may choose $\Psi_{grow}= 0$.
The reheating of the universe imposes a further constraint. There
should be a global minimum in the potential at $U_{min} = 0$, if
reheating is to proceed through coherent oscillations of the
inflaton. The region of the potential that drove the inflationary
expansion is then constrained by cosmological observations, as in
the standard scenario.

\begin{figure}[h]
\begin{center}
\includegraphics[width=2.7in,angle=0,clip=true]{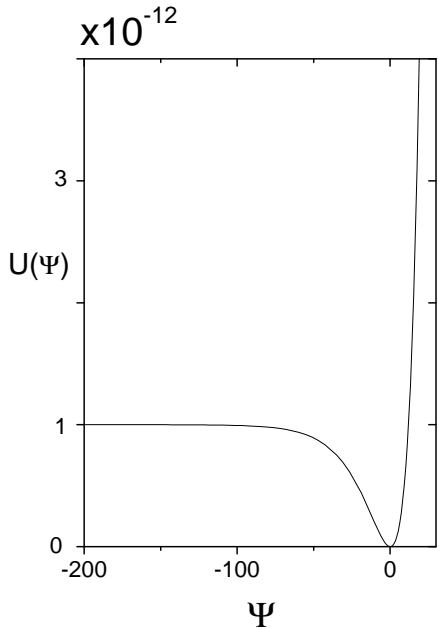}
\caption{The plot shows an emergent scalar potential that allows
for conventional reheating.} \label{pot}
\end{center}
\end{figure}

Motivated by the former discussion, we consider the following
potential as an example:

\begin{equation}\label{U}
U(\Psi)= U_0 \left[\exp\left(\beta
\Psi/\sqrt{3}\right)-1\right]^2,
\end{equation}
which exhibits the generic properties described above, see
Fig.~(\ref{pot}).

Potentials of this form have been considered previously, not only
in the context of the emergent universe \cite{Ellis:2003qz,
Ellis:2002we, Mulryne:2005ef} but  in a number of different
settings, including cases that introduce higher-order curvature
invariants  into the Einstein-Hilbert action. Such corrections are
required when attempting to renormalize theories of quantum
gravity \cite{Antoniadis:1986tu}. They also arise in low-energy
limits of superstring theories \cite{Candelas:1985en}. In general,
these theories are conformally equivalent to Einstein gravity plus
a minimally coupled, self-interacting scalar field. In particular,
potentials with the structure of Eq.~(\ref{U}) can be obtained
from theories that include a $R^2$ term in the action, where $R$
is the Ricci scalar \cite{Ellis:2003qz}. In general, all these
potentials possess a global minimum at $U = 0$.
Following Ref.~\cite{Mulryne:2005ef}, we take $U_0 = 10^{-12}$ and
$\beta = 0.1$ as representing typical values satisfying the
constraints imposed by the WMAP satellite \cite{Peiris:2003ff,
Spergel:2003cb}.

As an example of a Brans-Dicke potential that satisfies the
condition of static solution, Eqs.~(\ref{cc1}) and (\ref{cc2}), we
consider the following polynomial potential

\begin{equation}\label{JBDpot}
V(\Phi) = V_0 + A\left(\Phi - \Phi_0 \right)
+\frac{1}{2}\,B\left(\Phi - \Phi_0 \right)^2   +
\frac{1}{4!}\,\,C\left(\Phi - \Phi_0 \right)^4\,,
\end{equation}
where $\Phi_0$ correspond to the value of the JBD field at the
static solution.

Following the discussion of Sect.~\ref{inf}, we choose the
parameter $C$ in order to force $\Phi$ to settle down to a
non-zero expectation value, $\Phi \rightarrow \Phi_f$. Then we
have:

\[
%
C=-12\left[\frac{2V_0 - 2A\,\Phi_0 + B\,\Phi^2_0 + A\,\Phi_f -
B\,\Phi_0\,\Phi_f}{(\Phi_0 - \Phi_f)^3(\Phi_0 + \Phi_f)}\right].
\]

The parameters $V_0, A, B$ are fixed in order to obtain a static
solution at $\Phi=\Phi_0$, $a=a_0$ and $\rho=\rho_0$

\begin{eqnarray}
V_0 &=& \frac{3\Phi_0}{a_0}-\rho_0\,,\\
A &=& \frac{3}{a_0^2}\,,\\
B &=& \frac{\chi}{2\,a^2_0\,\Phi_0}\,,
\end{eqnarray}
where we have introduced the dimensionless parameter $\chi$,
satisfying  $0< \chi <3/2$. The JBD parameter requires:

\begin{eqnarray}
\label{w}
-\frac{3}{2} < & w & <-\frac{\sqrt{3}}{4}\,\sqrt{3-2\,\chi} -
\frac{1}{4}(3+\chi)\,.
\end{eqnarray}

\begin{figure}[h]
\begin{center}
\includegraphics[width=1.5in,angle=0,clip=true]{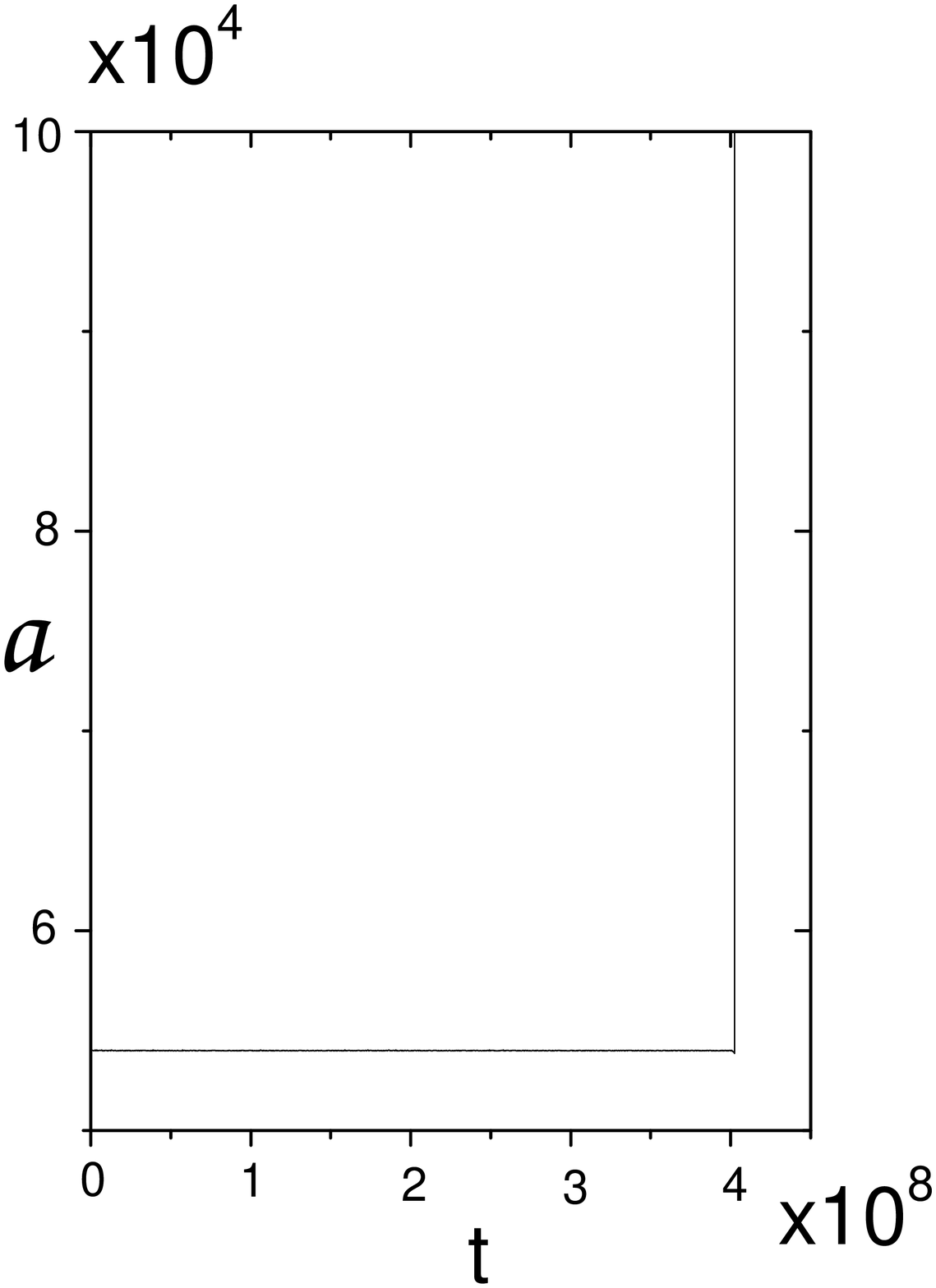}
\includegraphics[width=1.5in,angle=0,clip=true]{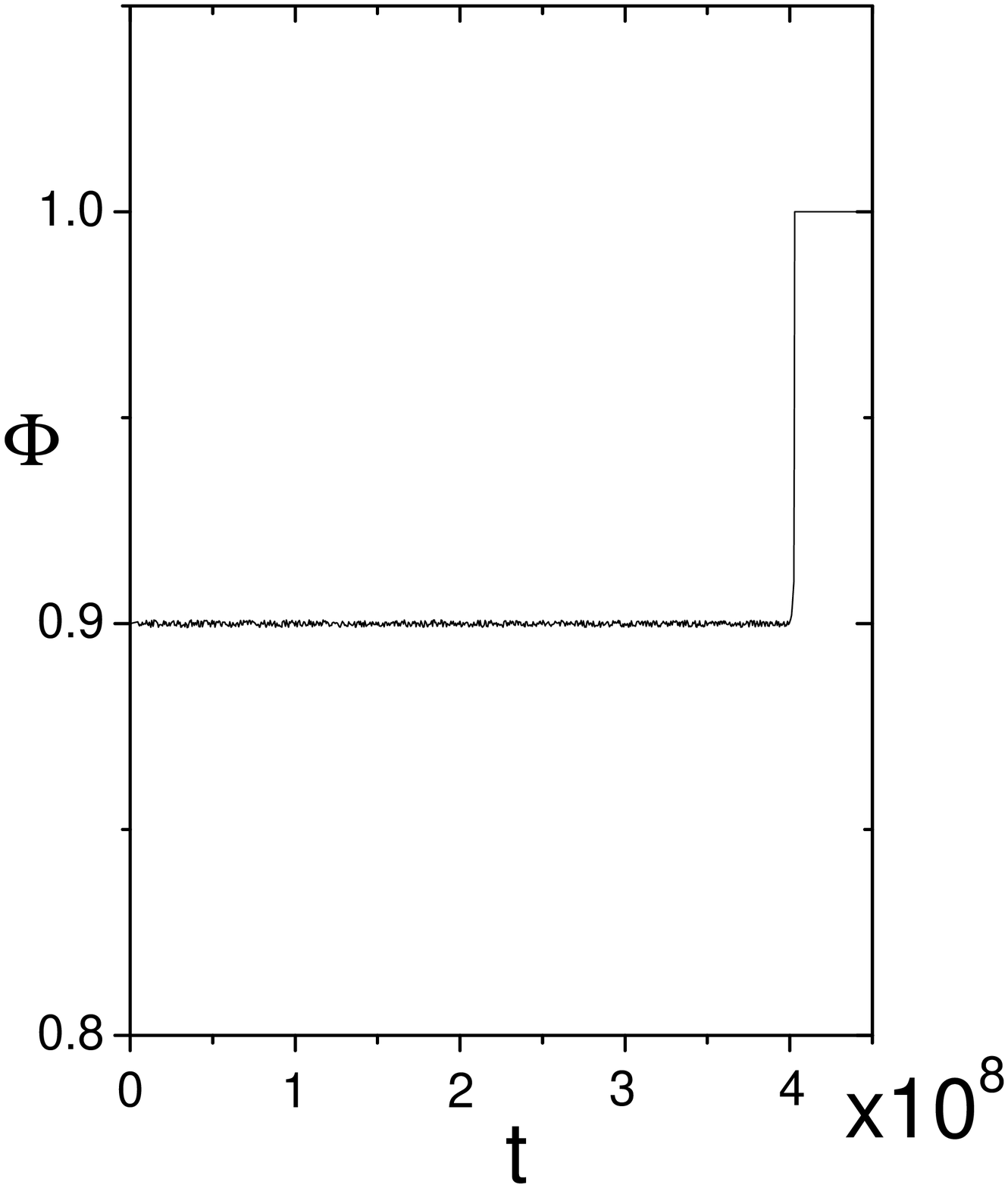}
\includegraphics[width=1.5in,angle=0,clip=true]{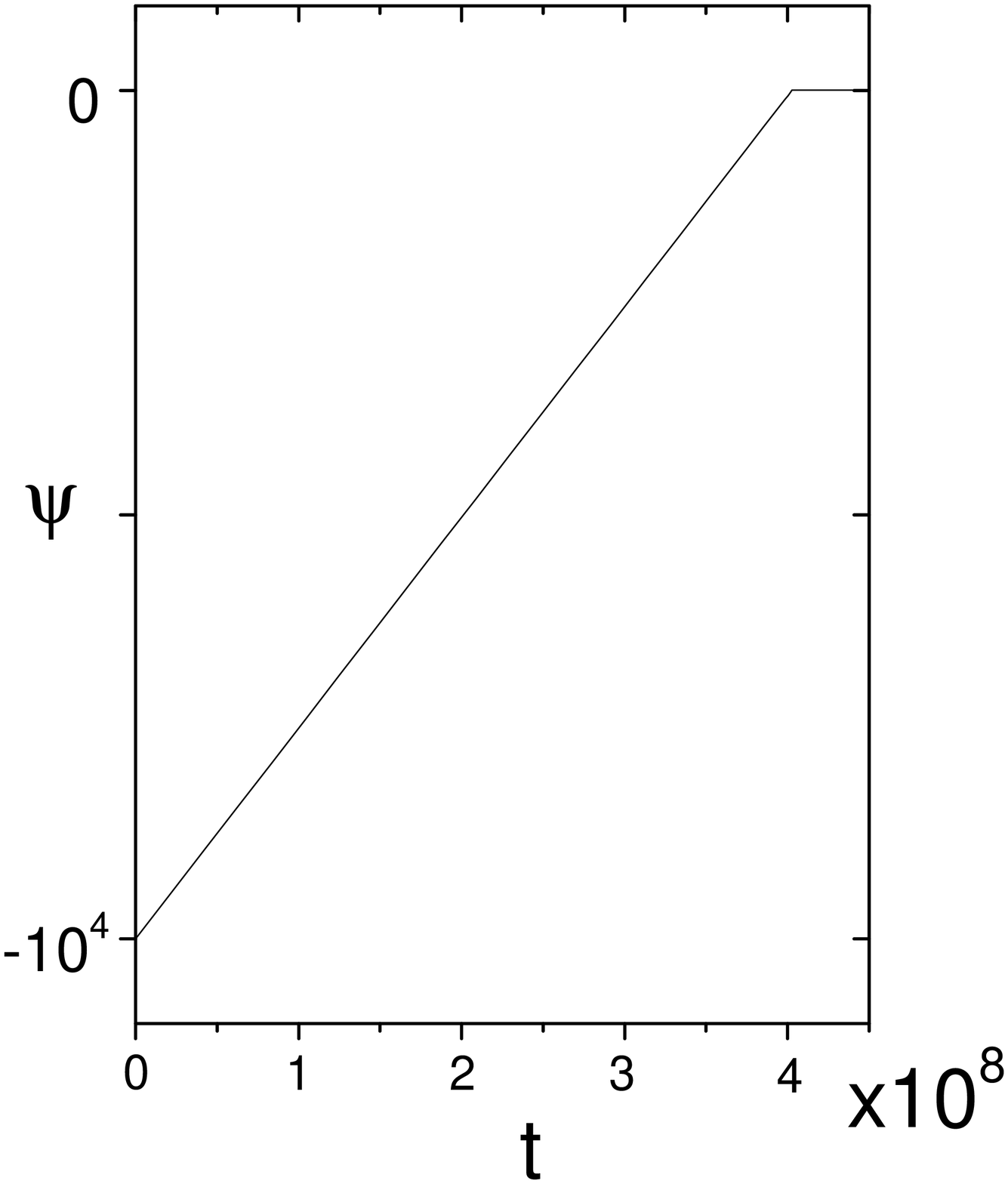}
\caption{Behavior of the scale factor $a$, the JBD field $\Phi$
and the scalar field $\Psi$ as function of the cosmological time
$t$.} \label{sol}
\end{center}
\end{figure}

The static energy density is given by:

\begin{equation}
\rho_0 = \frac{\Phi_0}{a_0^2} + U_0\,.
\end{equation}

In order to obtain a numerical solution we take the following
values for the parameters in the JBD potential $\Phi_0 =0.9$,
$\Phi_f =1$, $a_0 = 5.4\times 10^4$, $\chi =1$ and $w = -1.45$,
where we have used units  in which $8\pi G=1$. These particular
parameters satisfy all the constraints discussed previously. On
the other hand, in order to consider the model just at the
classical level we have to be out of the Planck era. This imposes
the following conditions:  $\rho < \Phi(t)^2$ and $V(\Phi) <
\Phi(t)^2$, which are satisfied with the values of the parameters
mentioned above.

Now, let us consider a numerical solution corresponding to a
universe starting from an initial state close to the static
solution. The whole evolution of the model is shown in
Fig.~(\ref{sol}), where we can notice that during the part of the
process where the scalar field $\Psi$ moves in the asymptotically
flat region of the potential $U(\Psi)$, the universe remains
static, due to the form of the JBD potential, Eq.~(\ref{JBDpot}).
This means that the scale factor and the JBD field do not evolve
with the cosmological time keeping their equilibrium values $a_0$
and $\Phi_0$ during this period.

\begin{figure}[h]
\begin{center}
\includegraphics[width=1.5in,angle=0,clip=true]{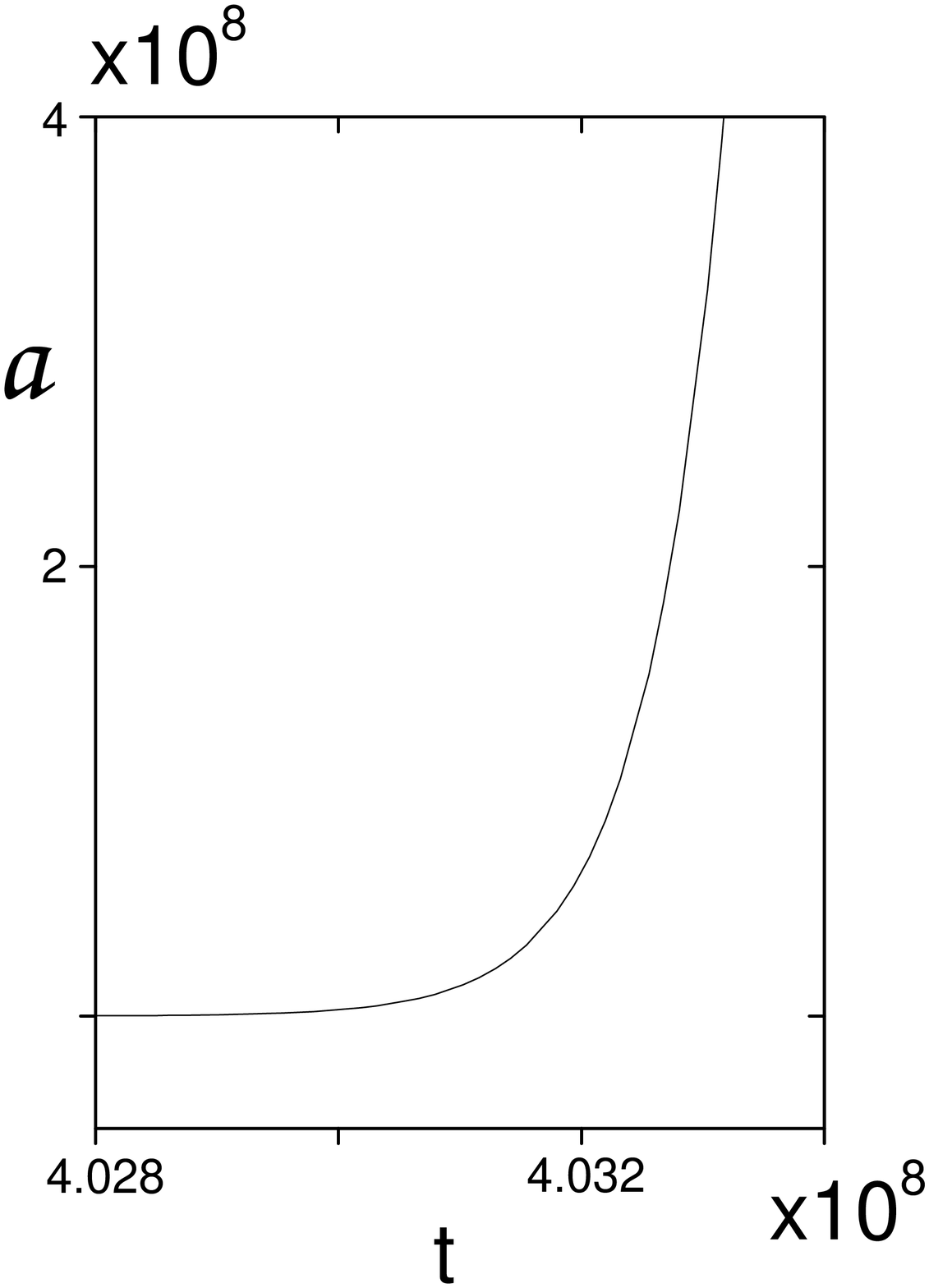}
\includegraphics[width=1.5in,angle=0,clip=true]{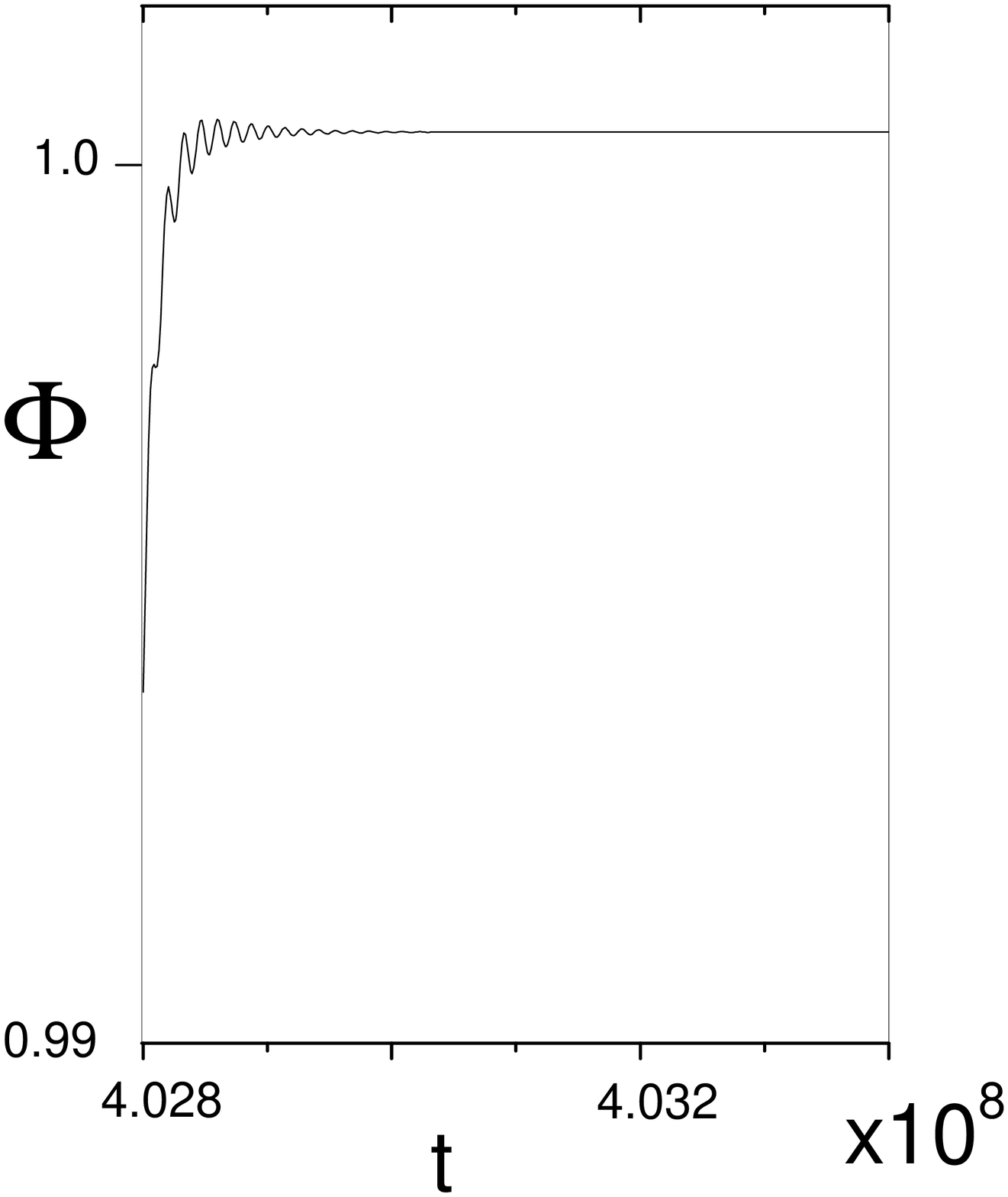}
\includegraphics[width=1.5in,angle=0,clip=true]{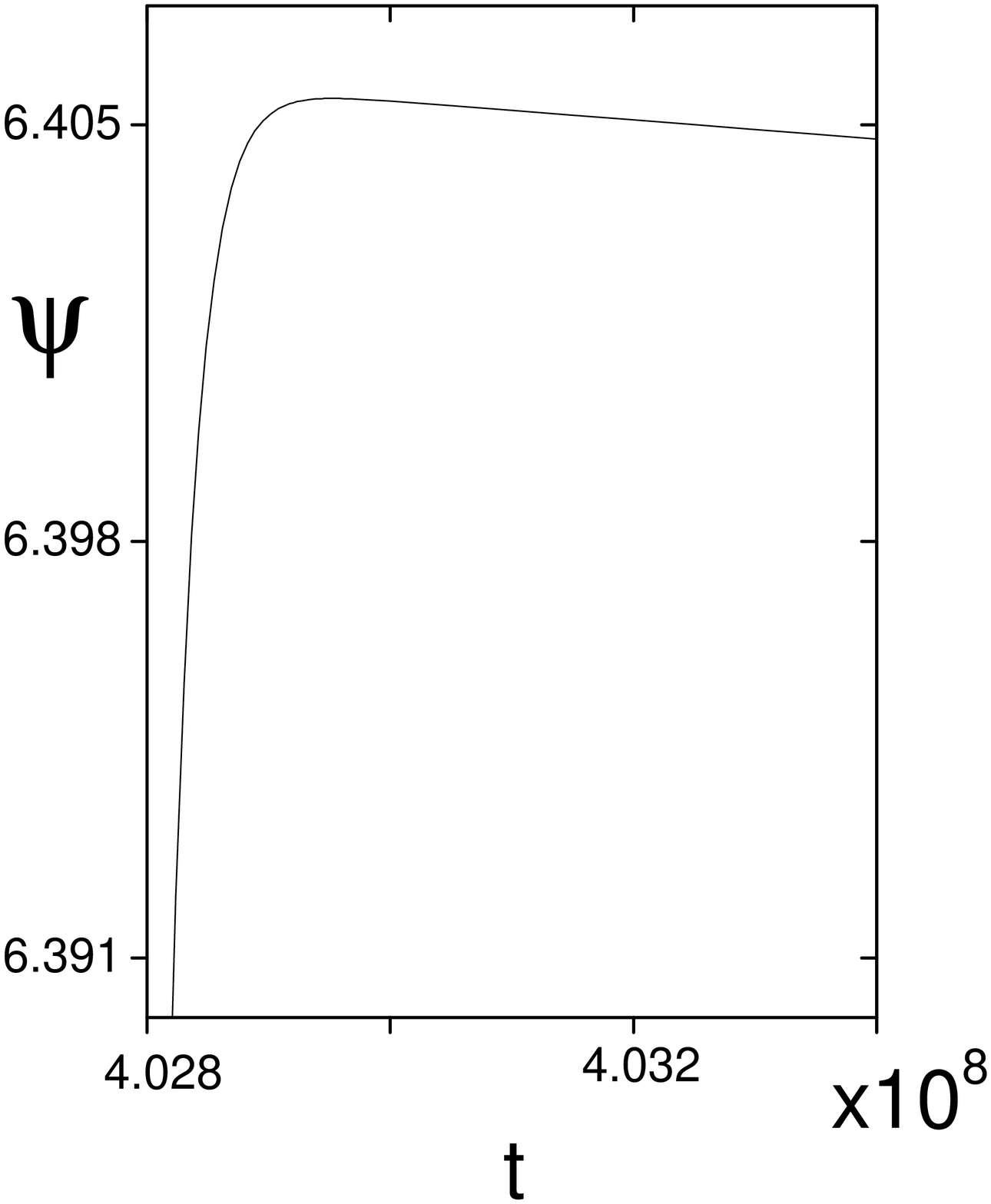}
\caption{Plot showing $a$, $\Phi$ and $\Psi$ near the beginning of
the inflationary period.}\label{infl}
\end{center}
\end{figure}

The static regimen finishes when the scalar field moves past the
minimum of its potential (near the value $\Psi\sim 0$) and begins
to decelerate  as it moves up its potential. During this period
the static equilibrium is broken, and the scale factor and with
the JBD field start to evolve. Finally, the scalar field starts to
go down the potential $U(\Psi)$ in the slow-roll regimen.
The details of the last part of this process is shown in
Fig~(\ref{infl}), where we  note that at the moment when the
scalar field starts to roll back down its potential, the JBD field
attains  its  quasi-static value $\Phi \sim 1$, and the scale
factor starts its inflationary expansion.

\begin{figure}[h]
\begin{center}
\includegraphics[width=2.5in, angle=0,clip=true]{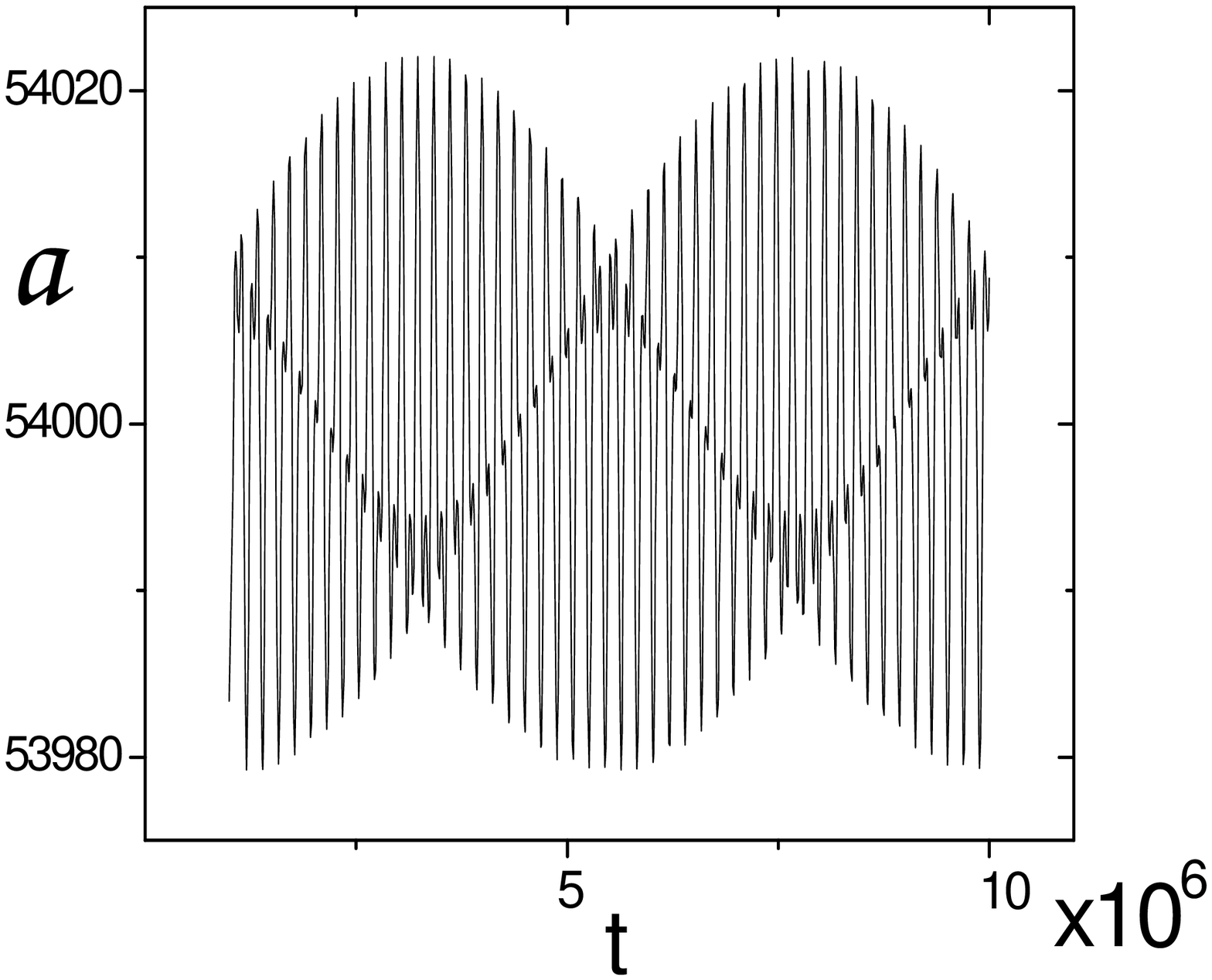}
\includegraphics[width=2.501in,angle=0,clip=true]{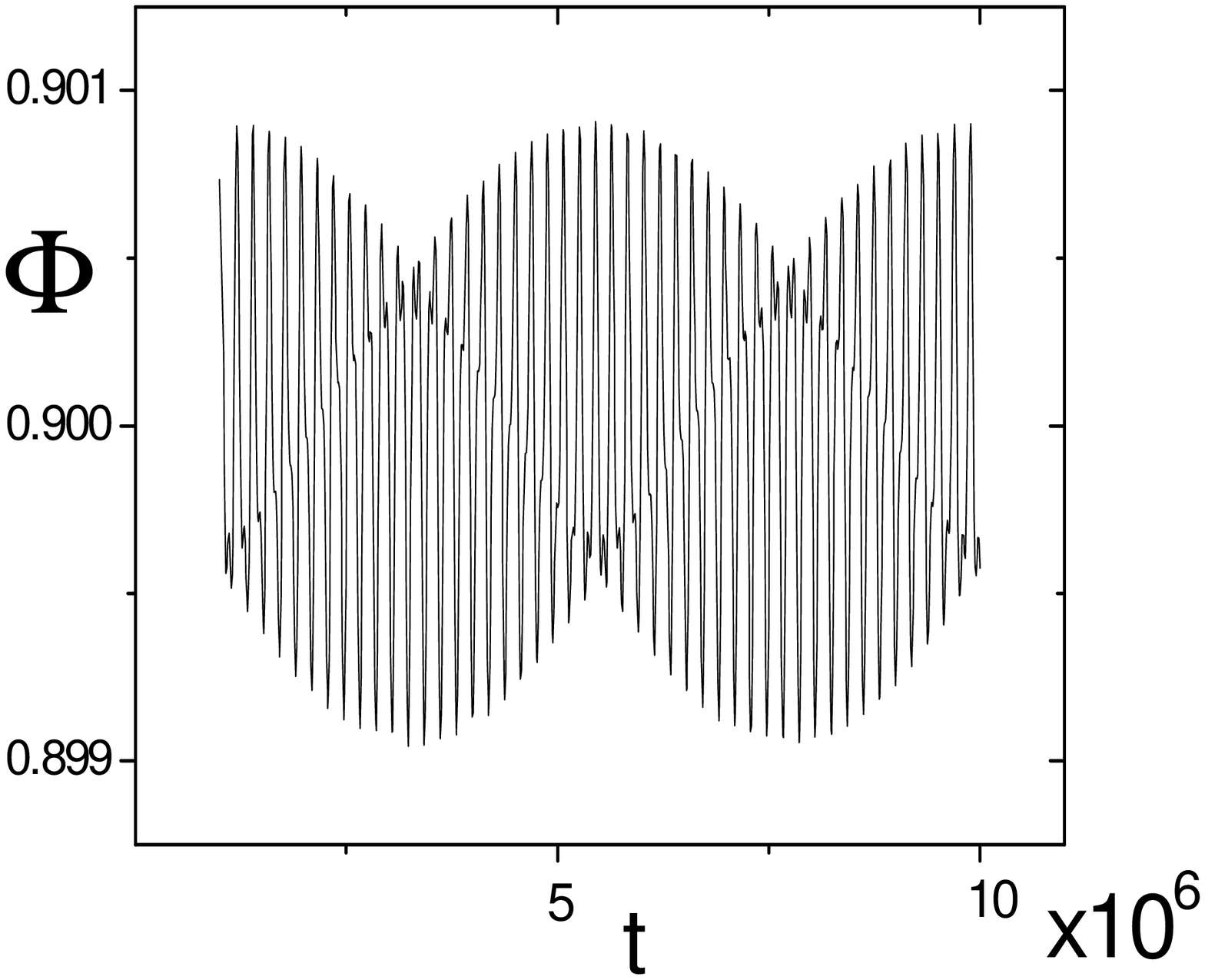}
\caption{Graph showing the scale factor and the JBD field
oscillating around their equilibrium values $a_0$ and $\Phi_0$,
while the scalar field follows  the constant
potential.}\label{osc}
\end{center}
\end{figure}

On the other hand, a numerical solution corresponding to a
universe starting from an initial state not in the static solution
but close to it, presents small oscillations around the
equilibrium values, as shown in Fig~(\ref{osc}). This tells us
that the static solution is stable.

\section{Conclusions}
\label{conc}

In this paper we have studied a Jordan-Brans-Dicke model and we
have determined whether that model could display the general
characteristics required for an emergent universe scenario. That
is a stable static past asymptotic solution followed by a period
of de Sitter inflation.

The original idea of an "emergent universe"~\cite{Ellis:2002we} is
a simple closed inflationary model in which the universe emerges
from an Einstein static state with radius $a_0 >> L_p$, inflates
and is then subsumed into a hot Big Bang era. The attractiveness
of the proposed model is that one can avoid an initial
quantum-gravity stage if the static radius is larger than the
Planck length.
However, this model suffers from the problem of instability of the
Einstein static state (see
Refs.~\cite{Mulryne:2005ef,Banerjee:2007qi,Nunes:2005ra,Lidsey:2006md})
which makes it extremely difficult to maintain its state for an
infinitely long time in the presence of fluctuations, such as
quantum fluctuations, thereby aborting the scenario.

In this work, we have provided an explicit construction of an
emergent universe scenario, which presents a stable past eternal
static solution and brings us the possibility of avoiding  an
initial quantum-gravity stage  if we chose the static radius to be
larger than the Planck length.

In particular, we have considered a JBD theory with a self
interacting potential and matter content corresponding to a scalar
field.

In the first part of the paper, we studied static solutions. In
order to do so we  determined the characteristics of the JBD
potential and we took  a constant scalar matter potential.
In particular, we have found a static solution in which the JBD
potential had a non-zero value and a positive derivative at the
static point $\Phi=\Phi_0$. In determining the stability of this
solution, we have calculated the real frequencies of small
oscillation about the static solution. This imposed a bound for
the first and second derivative of the JBD potential at the static
point $0< V''_0 < V'_0/(2\Phi_0)$. A restriction on the value of
the JBD parameter $w$ was also obtained (see Eq.~(\ref{cc2})).
In this way,  we have shown that it is possible to obtain a past
eternal universe in a JBD model, depending upon the
characteristics of the JBD potential and the Brans-Dicke parameter
$w$.

In the second part of the paper,  we studied the possibility that
our model present a past eternal static solution, which eventually
enters a phase where the stability of the static solution is
broken by changing the matter scalar field potential, thereby
leading  to a phase of inflation.

In our model, the mechanism that enables the universe to emerge
depends on the form of both potentials: the JBD potential
$V(\phi)$ and the matter scalar field potential, $U(\Psi)$. For
the scalar field potential it is required that it asymptotically
approaches   a constant value as $\Psi \rightarrow -\infty$, and
in order to break the cycles it should grow in magnitude for
larger $\Psi$. The JBD potential must satisfy similar requirements
to that described in Ref.~\cite{Green:1996xe}.

In the third part of the paper, we studied  a particular matter
scalar potential,  similar to the one used in the context of
emergent universe \cite{Ellis:2003qz, Ellis:2002we,
Mulryne:2005ef},  and a polynomial JBD potential,  which satisfies
the requirement of static stable past eternal solution followed by
a period of inflation, Eq~(\ref{condinf}).

We obtained numerical solutions for a universe starting from an
initial state close to the static solution. The numerical
solutions showed a behavior just like  that discussed in previous
sections. In particular we have found that when the scalar field
$\Psi$ moves in the asymptotically flat region of the potential
$U(\Psi)$, the scale factor and the JBD field experience small
oscillations about their equilibrium points. After that, when the
scalar field passes  the minimum of its potential and begins to
decelerate as it moves up its potential, we  found that the static
equilibrium is broken, and the scale factor and  JBD field start
to evolve. In particular, the numerical solution shows that when
the scalar field starts to go down the potential $U(\Psi)$, the
JBD field gets its quasi-static value and the scale factor begins
a quasi-exponential expansion.

We should note that a more detailed analysis of this process could
be done by using a dynamical system approach. We expect to return
to this point in the near future.

\begin{acknowledgments}
One of the authors (S. del C.) thanks Alex Vilenkin for reading
and comments about the manuscript. P.L. thanks the Institute of
Cosmology at Tufts University for its warm hospitality. S. del C.
is supported by the COMISION NACIONAL DE CIENCIAS Y TECNOLOGIA
through FONDECYT Grant N$^{0}$s. 1070306, 1051086 and 1040624, and
also was partially supported by PUCV Grant N$^0$. 123.787/2007. R.
H. is supported by the ``Programa Bicentenario de Ciencia y
Tecnolog\'{\i}a" through the Grant ``Inserci\'on de Investigadores
Postdoctorales en la Academia" \mbox {N$^0$. PSD/06}. P. L. is
supported by the COMISION NACIONAL DE CIENCIAS Y TECNOLOGIA
through FONDECYT Postdoctoral Grant N$^0$. 3060114.

\end{acknowledgments}



\begin{thebibliography}{99}



\bibitem{Guth}  Guth A.,  The inflationary universe: A possible solution to the horizon and flatness
problems, 1981 Phys. Rev. D {\bf 23} 347 .

\bibitem{Albrecht}  Albrecht A. and  Steinhardt P. J.,
 Cosmology for grand unified theories with radiatively induced symmetry
 breaking, 1982 Phys. Rev. Lett. {\bf 48} 1220.

\bibitem{Linde1}  Linde A. D.,  A new inflationary universe scenario: A possible solution of the horizon,
flatness, homogeneity, isotropy and primordial monopole problems,
1982 Phys. Lett. {\bf 108B} 389.

\bibitem{Linde2}  Linde A. D.,  Chaotic inflation, 1983 Phys. Lett. {\bf 129B} 177.

\bibitem{libro}  Linde A. D., \emph{Particle physics and inflationary
cosmology}, arXiv:hep-th/0503203 (Harwood Academic Publishers,
1990).


\bibitem{Borde:1993xh}
  Borde A. and Vilenkin  A.,
  Eternal inflation and the initial singularity,
 1994 Phys.\ Rev.\ Lett.\  {\bf 72} 3305.

\bibitem{Borde:1997pp}
  Borde A. and Vilenkin A.,
  Violation of the weak energy condition in inflating spacetimes, 1997
  Phys.\ Rev.\  D {\bf 56} 717.

\bibitem{Guth:1999rh}
 Guth A.~H.,
  Eternal inflation,
  arXiv:astro-ph/0101507.

\bibitem{Borde:2001nh}
  ~Borde A., ~Guth A.~H. and ~Vilenkin A.,
  Inflationary space-times are incompletein past directions, 2003
  Phys.\ Rev.\ Lett.\  {\bf 90} 151301.


\bibitem{Vilenkin:2002ev}
  ~Vilenkin A.,
  Quantum cosmology and eternal inflation,
  arXiv:gr-qc/0204061.



\bibitem{Ellis:2002we}
 ~Ellis G.~F.~R. and ~Maartens R.,
  The emergent universe: Inflationary cosmology with no singularity, 2004
  Class.\ Quant.\ Grav.\  {\bf 21} 223.

\bibitem{Ellis:2003qz}
  ~Ellis G.~F.~R., ~Murugan J. and~Tsagas C.~G.,
  The emergent universe: An explicit construction, 2004
  Class.\ Quant.\ Grav.\  {\bf 21} 233.

\bibitem{Mulryne:2005ef}
 ~Mulryne D.~J., ~Tavakol R., ~Lidsey J.~E. and ~Ellis G.~F.~R.,
  An emergent universe from a loop, 2005
  Phys.\ Rev.\  D {\bf 71} 123512.


\bibitem{Mukherjee:2005zt}
~Mukherjee S., Paul B.~C., Maharaj S.~D.  and ~Beesham A.,
  Emergent universe in Starobinsky model,
  arXiv:gr-qc/0505103.
%

\bibitem{Mukherjee:2006ds}
~Mukherjee S., Paul B.~C., Dadhich N.~K., Maharaj S.~D. and
Beesham A.~,
  Emergent universe with exotic matter, 2006
  Class.\ Quant.\ Grav.\  {\bf 23} 6927.

\bibitem{Banerjee:2007qi}
~Banerjee A., Bandyopadhyay T. and ~Chakraborty S.,
  Emergent universe in brane world scenario,
  arXiv:0705.3933 [gr-qc].

\bibitem{Nunes:2005ra}
 ~Nunes N.~J.,
  Inflation: A graceful entrance from loop quantum cosmology, 2005
  Phys.\ Rev.\  D {\bf 72} 103510.

\bibitem{Lidsey:2006md}
~Lidsey J.~E. and~Mulryne D.~J.,
  A graceful entrance to braneworld inflation, 2006
  Phys.\ Rev.\  D {\bf 73} 083508.



\bibitem{Gibbons:1987jt}
  ~Gibbons G.~W.,
  The entropy and stability of the universe, 1987
  Nucl.\ Phys.\  B {\bf 292} 784 .

\bibitem{Gibbons:1988bm}
 ~Gibbons G.~W.,
   Sobolev's inequality, Jensen's theorem and the mass and entropy of the
  universe, 1988
  Nucl.\ Phys.\  B {\bf 310} 636.


\bibitem{Jbd}  Jordan P.,  The present state of Dirac's cosmological
hypothesis, 1959 Z.Phys. {\bf 157} 112;
  Brans C.~ and ~Dicke R.~H.,
  Mach's principle and a relativistic theory of gravitation, 1961
  Phys.\ Rev.\  {\bf 124} 925.





\bibitem{Freund:1982pg}
  ~Freund P.~G.~O.,
  Kaluza-Klein cosmologies, 1982
  Nucl.\ Phys.\  B {\bf 209} 146.

\bibitem{Appelquist:1987nr}
  Appelquist T., Chodos A. and ~Freund P.~G.~O.,
%
\textit{Modern Kaluza-Klein theories} ( 1987 Addison-Wesley,
Redwood City).
%

\bibitem{Fradkin:1984pq}
  ~Fradkin E.~S. and Tseytlin A.~A.,
  Effective field theory from quantized strings, 1985
  Phys.\ Lett.\  B {\bf 158} 316.

\bibitem{Fradkin:1985ys}
 ~Fradkin E.~S. and ~Tseytlin A.~A.,
  Quantum string theory effective action, 1985
  Nucl.\ Phys.\  B {\bf 261} 1.

\bibitem{Callan:1985ia}
  Callan C.~G., Martinec E.~J., Perry M.~J.~ and ~Friedan D.,
  Strings in background fields, 1985
  Nucl.\ Phys.\  B {\bf 262} 593.

\bibitem{Callan:1986jb}
  CallanC.~G., Klebanov I.~R.~ and Perry M.~J.~,
  String theory effective actions, 1986
  Nucl.\ Phys.\  B {\bf 278} 78.

\bibitem{Green:1987sp}
  ~Green M.~B., ~Schwarz J.~H. and ~Witten E., \textit{Superstring theory}
 (1987 Cambridge, Uk: Univ. Pr., Cambridge Monographs On
Mathematical Physics).


\bibitem{La:1989pn}
 ~La D., ~Steinhardt P.~J. and Bertschinger E.~W.~,
  Prescription for successful extended inflation, 1989
  Phys.\ Lett.\  B {\bf 231} 231 .

\bibitem{Bertolami:1999dp}
 Bertolami O.~ and Martins P.~J.~,
  Non-minimal coupling and quintessence, 2000
  Phys.\ Rev.\  D {\bf 61} 064007.



\bibitem{Banerjee:2000mj}
  ~Banerjee N. and ~Pavon D.,
  Cosmic acceleration without quintessence, 2001
  Phys.\ Rev.\  D {\bf 63} 043504.

\bibitem{Sen:2001ki}
  Sen A.~A. and Sen S.,
Cosmology in scalar tensor theory and asymptotically de-Sitter
universe, 2001  Mod.\ Phys.\ Lett.\  A {\bf 16} 1303.

\bibitem{Green:1996xe}
  ~Green A.~M. and ~Liddle A.~R.,
  Open inflationary universes in the induced gravity theory, 1997
  Phys.\ Rev.\  D {\bf 55} 609.



\bibitem{Antoniadis:1986tu}
  ~Antoniadis I. and ~Tomboulis E.~T.,
  Gauge invariance and unitarity in higher derivative quantum gravity, 1986
  Phys.\ Rev.\  D {\bf 33} 2756.

\bibitem{Candelas:1985en}
  ~Candelas P., Horowitz G.~T., Strominger A. and ~Witten E.,
  Vacuum configurations for superstrings,  1985
  Nucl.\ Phys.\  B {\bf 258} 46 .



\bibitem{Peiris:2003ff}
  ~Peiris H.~V. {\it et al.}  [WMAP Collaboration],
  First year Wilkinson Microwave Anisotropy Probe (WMAP) observations:
  Implications for inflation, 2003
  Astrophys.\ J.\ Suppl.\  {\bf 148} 213.

\bibitem{Spergel:2003cb}
 ~Spergel D.~N. {\it et al.}  [WMAP Collaboration],
  First Year Wilkinson Microwave Anisotropy Probe (WMAP) Observations:
  Determination of cosmological parameters, 2003
  Astrophys.\ J.\ Suppl.\  {\bf 148} 175.





\end{thebibliography}
\end{document}